# Experimental signatures of the mixed axial-gravitational anomaly in the Weyl semimetal NbP

Johannes Gooth[1,2,*], Anna Corinna Niemann[1,3], Tobias Meng[4], Adolfo G. Grushin[5], Karl Landsteiner[6], Bernd Gotsmann[2], Fabian Menges[2], Marcus Schmidt[7], Chandra Shekhar[7], Vicky Süß[7], Ruben Hühne[3], Bernd Rellinghaus[3], Claudia Felser[7], Binghai Yan[7,8], Kornelius Nielsch[1,3]

[1] Institute of Nanostructure and Solid State Physics, Universität Hamburg, Jungiusstraße 11, 20355 Hamburg, Germany.

[2] IBM Research - Zurich, Säumerstrasse 4, 8803 Rüschlikon, Switzerland.

[3] Leibniz Institute for Solid State and Materials Research Dresden, Institute for Metallic Materials, Helmholtzstraße 20, 01069 Dresden, Germany.

[4] Institute for Theoretical Physics, Technical University Dresden, Zellescher Weg 17, 01062 Dresden, Germany.

[5] Department of Physics, University of California, Berkeley, CA 94720, USA.

[6] Instituto de Física Teórica UAM/CSIC, C/ Nicolás Cabrera 13-15, Universidad Autónoma de Madrid, Cantoblanco, 28049 Madrid, Spain

[7] Max Planck Institute for Chemical Physics of Solids, Nöthnitzer Straße 40, 01187 Dresden, Germany.

[8] Max Planck Institute for Physics of Complex Systems, Nöthnitzer Straße 38, 01187 Dresden, Germany.

* Correspondence to: jog@zurich.ibm.com.

**Abstract:** Weyl semimetals are materials where electrons behave effectively as a kind of massless relativistic particles known as Weyl fermions. These particles occur in two flavours, or chiralities, and are subject to quantum anomalies, the breaking of a conservation law by quantum fluctuations. For instance, the number of Weyl fermions of each chirality is not independently conserved in parallel electric and magnetic field, a phenomenon known as the chiral anomaly. In addition, an underlying curved spacetime provides a distinct contribution to a chiral imbalance, an effect known as the mixed axial-gravitational anomaly, which remains experimentally elusive. However, the presence of a mixed gauge-gravitational anomaly has recently been tied to thermoelectrical transport in a magnetic field, even in flat spacetime, opening the door to experimentally probe such type of anomalies in Weyl semimetals. Using a temperature gradient, we experimentally observe a positive longitudinal magneto-thermoelectric conductance (PMTC) in the Weyl semimetal NbP for collinear temperature gradients and magnetic fields ($\nabla T \parallel$ B) that vanishes in the ultra quantum limit. This observation is consistent with the presence of a mixed axial-gravitational anomaly. Our work provides clear experimental evidence for the existence of a mixed axial-gravitational anomaly of Weyl fermions, an outstanding theoretical concept that has so far eluded experimental detection.

**Main Text:** The conservation laws of physics, such as of charge, energy and momentum, play a central role in physics. In some special cases, classical conservation laws are broken at the quantum level by quantum fluctuations, in which case the theory is said to have quantum anomalies (1). One of the most prominent examples is the chiral anomaly (2,3), involving massless chiral fermions. These particles have their spin, or internal angular momentum, aligned either parallel or antiparallel with their linear momentum, labelled as left and right chirality respectively. In three spatial dimensions the chiral anomaly is the breakdown, due to externally applied parallel electric and magnetic fields (4), of the classical conservation law that dictates that the number of massless fermions of each chirality are seperately conserved. The current which measures the difference between left and right handed particles is called the axial current and is not conserved at the quantum level.

In condensed matter, massless chiral fermions exist in three-dimensional materials whose conductance and valence bands touch in isolated points. At energies in the vicinity of these points the electrons are effectively described by the Weyl Hamiltonian (5–9) which implies that the energy of these 'Weyl' fermions scales linearly with their momentum. Materials exhibiting such band touching points, known as Weyl nodes, are referred to as Weyl semimetals. Weyl nodes occur in pairs of opposite chirality (10) that are, in the absence of additional symmetries, topologically stable when they are separated in momentum space (Fig. 1A). The chiral Weyl fermions in the vicinity of these nodes are subject to a chiral anomaly, which results in a strong positive magneto-conductance (PMC) that can be detected experimentally (4, 11). Inspired by the pioneering studies of the chiral anomaly in pion physics (2, 3), several research groups have recently reported on the observation of chiral-anomaly-induced longitudinal PMC in $Na_3Bi$ (12), TaAs(13), NbP(14), GdPtBi(15), $Cd_2As_3$(16), TaP(17) and RPtBi(18).

Three-dimensional chiral fermions are theoretically predicted to also exhibit a mixed axial-gravitational anomaly (19–21). In curved spacetime, the anomaly contributes to the violation of the covariant conservation laws of the axial current, relevant for the chiral anomaly, and to the conservation law for the energy-momentum tensor (22). The energy-momentum tensor encodes the density and flux of energy and momentum of a system. The mixed axial-gravitational anomaly has been suggested to be relevant to the hydrodynamic description of neutron stars (23), and the chiral vortical effect in the context of quark-gluon plasmas (24). However, a clear experimental signature so far remains elusive.

While the flatness of space time would imply that gravitational anomalies are irrelevant for condensed matter systems, it has been recently understood that the presence or absence of a positive magneto thermoelectric conductivity (PMTC) of Weyl fermions is tied to the presence or absence of a mixed axial-gravitational anomaly in flat spacetime (24–26). In the SI, we demonstrate the connection of the mixed axial-gravitational anomaly and the observed PMTC by a calculation merely based on the conservation laws for charge and energy, the standard Kubo formalism for the conductivities, and a relaxation time approximation. Since the Weyl semimetal lives in a flat spacetime, the mixed axial-gravitational anomaly does not directly affect the conservation laws for charge and energy. An anomalous contribution to the energy current has nevertheless been identified in the Kubo formalism (22). Inserting the latter into the conservation laws and using a simple relaxation time approximation, we find that thermoelectric transport in flat spacetime is explicitly modified due to the presence of the mixed axial-gravitational anomaly in the underlying field theory.

The connection of thermal transport and the mixed axial-gravitational anomaly is also apparent in a relativistic quantum field theory computation of transport properties (24), as well as the hydrodynamic formalism of the effective chiral electron liquid (25, 26). In the latter approach, the presence of a mixed axial-gravitational anomaly modifies the thermodynamic constitutive relations of the current and energy-momentum tensor in terms of gradients of the relevant hydrodynamic variables: temperature, chemical potential, and velocity (25). These modifications can be viewed as the hydrodynamic equivalent of the anomalous contributions to the energy current identified in the Kubo formalism. While the Kubo based calculation presented in the SI is thus quantitatively consistent with the hydrodynamic calculation of (26), we note that transport in current Weyl semimetal samples is not consistent with the hydrodynamic regime which involves strong interactions, and features fast energy-momentum relaxation between the nodes. The predicted PMTC is furthermore consistent with the semiclassical approach based on the Boltzmann equation (27–30), which so far lacks a simple connection to the anomalous origin of this contribution. This consistency of different theoretical approaches illustrates that anomalies impact transport on a fundamental level; their effect can consistently be derived based on any calculation that keeps track of conservation laws and symmetries, and that correctly captures the topological character of a Weyl node. It is worth to emphasise that the relation between mixed axial-gravitational anomalies and thermal transport discussed above is conceptually independent of the useful tool put forward by Luttinger (31) to compute thermal responses, which relies on the formal equivalence between responses to thermal and gravitational gradients.

The PMC and PMTC are fundamentally linked to the charge current response ($\mathbf{J}$) when an electric field $\mathbf{E}$ and a thermal gradient $\nabla T$ are applied through the relation, $\mathbf{J} = G\mathbf{E} + G_T \nabla T$. Here, $G = J_x/E_x$ denotes the electrical conductance along the $x$-direction, characterizing the electrical current response on electric fields, and $G_T = J_x/\partial_x T$ is the thermoelectrical conductance, characterizing the electrical current in the Weyl metal induced by a temperature gradient. In low magnetic fields, the mixed axial-gravitational anomaly and the chiral anomaly imply a positive magneto-current contributions to the transport coefficients $G = d_e + c_1 a_\chi^2 B_\parallel^2$ and $G_T = d_{\text{th}} + c_2 a_\chi a_g B_\parallel^2$ with $c_{1,2} > 0$. Here $d_e$ and $d_{\text{th}}$ express classical Drude parts and the coefficients $a_\chi$ and $a_g$ account for the contributions of the chiral and mixed axial-gravitational anomaly, respectively (26–28, 32, and SI). In the ultra quantum limit at high magnetic fields, when only the lowest Landau level contributes to transport, $G$ depends linearly on the magnetic field, and the gravitational anomaly does not contribute to $G_T$ (see below and SI). Analogous to the PMC, which requires a parallel electric and magnetic field as determined by the chiral anomaly, the PMTC is expected to be locked to the magnetic-field direction due to the anomalous contribution(26–30). The combined measurement of i) a finite value of $a_g$ ii) the functional dependence $\nabla T \cdot \mathbf{B} \neq 0$ of the PTMC at low fields, and iii) the absence of PTMC at high fields, is the experimental signature of the mixed axial-gravitational anomaly in thermal transport.

The magneto-thermoelectrical conductance of the half-Heusler GdPtBi (15) has recently been calculated from separate thermopower and electrical conductance measurements, obtaining a PMTC contribution at low B-fields. This PMTC was, however, interpreted as a signature of the node creation process, which depends on magnetic field. To obtain experimental signatures for the presence of the mixed axial-gravitational anomaly, it is therefore desirable to go beyond these experiments and investigate the electrical response of intrinsic Weyl semimetals to temperature gradients in collinear magnetic fields.

Here we report the observation of increased electrical magneto-currents in the Weyl semimetal NbP, separately biased by an electric field ($\mathbf{E} \parallel \mathbf{B}$) (Fig. 1B,C) and a temperature gradient ($\nabla T \parallel \mathbf{B}$) (Fig. 1D,E). In both cases, the currents are locked to the direction of the magnetic field, consistent with the predictions rooted in the chiral and the mixed axial-gravitational anomaly. In the ultra quantum limit at high magnetic fields, when $G$ depends linearly on the magnetic field, a vanishing of the anomaly-induced contribution to $G_T$ is observed.

For our experiments we used NbP micro-ribbons of $50\mu\text{m} \times 2.5\mu\text{m} \times 0.5\mu\text{m}$, cut out from single crystalline bulk samples with a Ga focused ion beam. The transport direction in our samples matches the [100] axis of the crystal (see SI for details). An on-chip micro-strip line heater near the micro-ribbon generates a temperature gradient across the ribbon's length, and relatively small temperature differences ($< 350\,\text{mK}$) ensure that the measurement is in the linear response regime (see Fig. S6 in SI). The temperature gradient $\nabla T$ was measured by resistance thermometry using two metal four-probe thermometer lines located at the NbP micro-ribbon ends as shown in Fig. 1D. The metal lines for thermometry also serve as electrodes for applying an electrical bias and for measuring the current response of the ribbon. The elongated geometry of the micro-ribbons with contact lines across the full width of the samples was chosen to ensure that current jetting is suppressed and to provide homogenous field distributions(15, 17). To justify the description of the carriers in terms of Weyl fermions it is essential that the Fermi level $E_F$ is as close as possible to the Weyl nodes of NbP (33, 34). By means of Ga doping, we recently showed that $E_F$ is located only 5 meV above the Weyl points, in the electron cone of our NbP sample (14).

In a first set of transport experiments of electrical conductance measurements under isothermal conditions ($\nabla T = 0$) we establish that the NbP micro-ribbon can be accurately described by Weyl fermions. For this purpose, a voltage $V = 1\,\mathrm{mV}$ is applied along the ribbon, which sets an electric field $\mathbf{E}$, and the corresponding current $\mathbf{J}$ is measured through a near-zero impedance ($1\,\Omega$) ammeter. When the magnetic field is switched on, the Weyl nodes split into Landau levels. For each Weyl node, the zeroth Landau level disperses linearly with momentum along the magnetic field $\mathbf{B}$ (Fig. 2A), and is thus chiral, unlike the remaining Landau levels, that disperse quadratically. Aligned electric and magnetic fields ($\mathbf{E} \parallel \mathbf{B}$) generate a chiral charge flow between the two valleys, with a rate that is proportional to $\mathbf{E} \cdot \mathbf{B}$ (4). To equilibrate the induced chiral imbalance between the left- and right-handed fermions, large-momentum internode scattering is required, which in general depends on $\mathbf{B}$ (11, 32). In the low-field regime, where many Landau levels are filled, it is possible to solve the corresponding Boltzmann (11) or hydrodynamic (26) transport equation which results in the positive chiral anomaly-induced magneto-conductance contribution $\Delta G = c_1 a_\chi^2 B_\parallel^2$. In the high-field limit, when only the lowest Landau levels contribute to transport, the magneto-conductance behaves linearly with applied field. This is the transport fingerprint of the chiral anomaly. As shown in Fig. 2B, we observe a large PMC up to room temperature for $\mathbf{E} \parallel \mathbf{B}$, which is sensitive to misalignments (Fig. 2C). While the low-field regime is well described by a quadratic fit ($\Delta G \sim B_\parallel^2$), in agreement with the Boltzmann description for chiral anomaly, the linear high-field regime can be explained by a transition from a multi-level state to the limit where only the lowest chiral Landau levels contribute to the transport (14). In accordance with chiral charge pumping, the PMC at low magnetic fields is well reflected by $\cos^2(\phi)$, where $\phi$ is the angle between **E** and **B** (Fig. 2D). The narrowing of the angular width at higher fields is caused by strong collimation of the axial beams (12). The observed locking pattern and the consistent quadratic low **B** dependence are the fundamental signatures of the chiral anomaly (11) and support the description of the system in terms of chiral Weyl fermions.

We now turn to testing the mixed axial-gravitational anomaly in the NbP micro-ribbon. We employ a transport experiment, but in this case applying a thermal gradient instead of a voltage bias. Importantly, because the NbP sample is shortcut through a near-zero impedance ammeter, no net electric field is imposed. Excluding electric fields is essential for a clear distinction from the chiral anomaly, which is induced by a finite **E** ∥ **B**. Instead, applying $\nabla T$ ∥ **B** leads to a net energy density difference between the two chiral valley fluids (22, 26), proportional to $\nabla T \cdot$ **B**, that is equilibrated through an intervalley energy transfer (Fig. 3A) The resulting imbalance leads to a charge current, through the chiral magnetic effect which then leads to the magneto-thermoelectrical conductance contribution $\Delta G_T = c_2 a_\chi a_g B_\parallel^2$ (22, 26–30). This allows us to probe the presence of the mixed axial-gravitational anomaly through its effect on thermoelectric transport in a condensed matter system. The data corresponding to these measurements is shown in Fig. 3B - C. The applied temperature gradient indeed appears to result in similar magneto-transport features as the application of an electric field. When $\nabla T$ and **B** are aligned in parallel, the thermoelectrical conductance at low magnetic fields exhibits a positive PMTC that fits to $G_T \sim B_\parallel^2$ with the same $\cos^2(\phi)$-locking pattern as the PMC (Fig. 3D). At high temperatures ($T \gtrsim 150K$), the observed dependence of the magneto-transport on the field strength is consistent with the presence of a mixed axial-gravitational anomaly and its corresponding thermoelectric transport prediction (26–30 and SI). At lower temperatures, however, we observe a decrease of $G_T$. This decrease occurs in the same magnetic field range than the crossover from a quadratic to a linear field dependence in $\Delta G$, in agreement with the fact that both effects can be explained by the crossover to a one-dimensional dispersion that Weyl metals show along **B** in the ultra quantum limit (15, 35). As we show in the SI, the suppression of thermoelectric transport at high magnetic fields occurs because $\Delta G_T$ is proportional to the derivative of the electron density with respect to temperature, $\Delta G_T \sim \partial \rho / \partial T$. Since the density of states is constant in the ultra quantum limit, $D(E) \approx D_0$, the electron density at large magnetic fields is independent of temperature, and $\Delta G_T = 0$.

The ratio of $G_T/G$ should exhibit another measurable transport coefficient, the thermopower $S$. Starting from the relation $\mathbf{J} = G\mathbf{E} + G_T \nabla T$, the thermopower can be determined either from a measurement using an open circuit ($\mathbf{J} = 0$), or from combining the above experiments at $\mathbf{E} = 0$ and $\nabla T = 0$. $S$ expresses the open-circuit voltage response to a temperature gradient. To carry out this test, we removed the shortcutting ammeter from our experiment and measured the open voltage response to a temperature gradient in a collinear magnetic field (Fig. 1A). As shown in Fig 4 B, the calculated $G_T/G$ matches the measured $S$ excellently. The agreement is an important cross-check, confirming the results and interpretation above. Furthermore, this result implies that the thermopower is suppressed not solely by the presence of the chiral anomaly as suggested in (15), but rather by the presence of both the chiral and mixed axial-gravitational anomalies.

In conclusion, our measurements reveal a positive longitudinal magneto-thermoelectric conductance (PMTC) in the Weyl semimetal NbP, a signature linked to the presence of the axial-gravitational anomaly of chiral fermions in three spatial dimensions. In short, the thermally biased experiment confirms the predicted $B_\parallel^2$-dependence of the thermoelectric conductance at low magnetic fields, its dependence on the relative orientation of the magnetic field and the thermal gradient, and the suppression of thermoelectric transport at high magnetic fields. These effects arise concurrently with the standard chiral anomaly, whose signatures we have observed in the field-induced correction to the standard electric conductance. Our results show that it is possible to detect the presence of the mixed axial-gravitational anomaly of Weyl fermions, particularly elusive in other contexts, in relatively simple transport experiments using a macroscopic condensed matter system in a flat spacetime.

**Figure 1  Positive magneto-conductance $G(\mathbf{B}_\parallel)$ and magneto-thermoelectric conductance $G_T(\mathbf{B}_\parallel)$ in the Weyl semimetal NbP.** (A) Sketch of two Weyl cones with distinct chiralities $+\chi$ and $-\chi$, represented in green and red, respectively. The two chiral nodes are separated by $2k_D$ in momentum space. (B) False-colored optical micrograph of the measurement device. The electrical conductance $G = J/V$ is measured under isothermal conditions ($\nabla T = 0$) along an NbP micro-ribbon (green) by applying a constant voltage $V$ and measuring the electrical current response $J$. (C) The magneto-thermoelectric conductance $G_T = J/T$ is measured on the same device by applying a temperature gradient $\nabla T$, but no electric field ($V = 0$). (D) Longitudinal $G(\mathbf{E} \parallel \mathbf{B})$ and (E) $G_T(\nabla T \parallel \mathbf{B})$ as a function of the magnetic field $\mathbf{B}$ at a base temperature of $25\,\mathrm{K}$ (solid lines). At low magnetic fields, $G(\mathbf{E} \parallel \mathbf{B})$ and $G_T(\nabla T \parallel \mathbf{B})$ follow a parabolic law (dotted lines), consistent with chiral and gravitational anomaly, respectively. The negative sign of $G_T$ accounts for electron transport.

**Figure 2  Chiral anomaly in NbP.** (A) In a strong magnetic field, the Weyl nodes quantize into Landau levels (LLs). The lowest LLs exhibit a linear dispersion with distinct chirality. Parallel electric and magnetic fields pump chiral charges from one cone into the other, which breaks chiral symmetry. (B) Longitudinal magneto-conductance (MC) without zero-field contributions $\Delta G$ at selected temperatures. The transition from a quadratic dependence at low $B_\parallel$ to a linear regime at high fields indicates a transition from the multi-subband to the fully quantized regime, where only the lowest LLs are occupied. (C) $\Delta G$ versus $|\mathbf{B}|$ for different angles $\phi$ between the electric and the magnetic field. The MC

changes from positive to negative sign for $\phi > 35°$ at high fields. (D) Angular dependence of the axial current. At $|\mathbf{B}| < 3\,\mathrm{T}$, the axial current is reasonably well described by a squared cosine function. However, at higher fields the angular width narrows considerably, indicating a strongly collimated axial current.

**Figure 3  Evidence of the mixed axial-gravitational anomaly in NbP.** (A) Parallel temperature gradients and magnetic fields result in a transfer of particles and energy ($E$) from one cone to the other. (B) Negative longitudinal magneto-thermoelectric conductance (MTC) without zero-field contributions $-\Delta G_T$ at selected temperatures. The negative sign accounts for electron transport. As expected for a mixed axial-gravitational anomaly, the MTC exhibits a quadratic low field-dependence. At higher fields, the formation of one-dimensional Weyl Landau levels dispersing only along the magnetic field direction strongly suppresses $G_T$ (c.f. SI). This effect weakens with increasing temperatures, which we ascribe to thermal excitation. (C) $\Delta G$ versus $|\mathbf{B}|$ for different angles $\phi$ between the electric and the magnetic field. The MC changes from positive to negative sign for $\phi > 30°$ at high fields. (D) Angular dependence of the axial current. Similar to the electrical conductance, the MTC at $|\mathbf{B}| < 3\,\mathrm{T}$ is reasonably well described by a squared cosine function. However, at higher fields, the angular width narrows considerably, indicating a strongly collimated energy flux.

**Figure 4  Longitudinal thermopower.** (A) False-colored optical micrograph of the thermopower measurement device. The thermopower $S = V/\Delta T$ is measured along a NbP

micro-ribbon (green) by applying a temperature difference $\Delta T$ and measuring the open-circuit voltage response $V$. (B) Measured longitudinal magneto-thermopower $S$ and calculated magneto-thermopower $G_T/G$ of the anomalous transport coefficients match quite well, as exemplarily shown for $25\,\text{K}$.

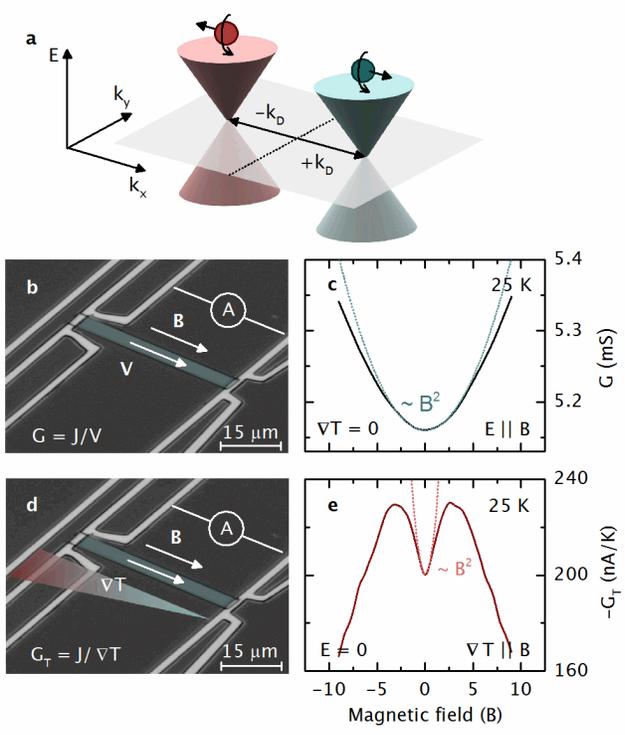

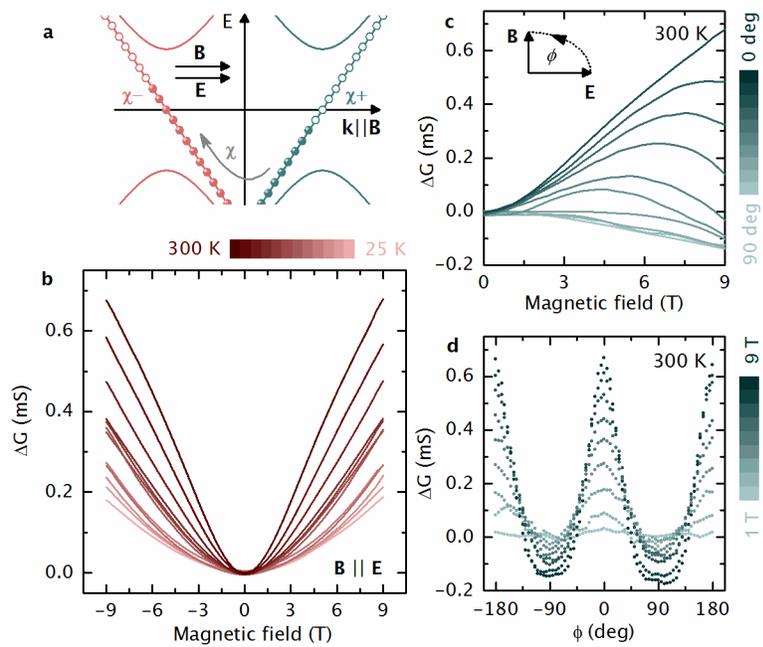

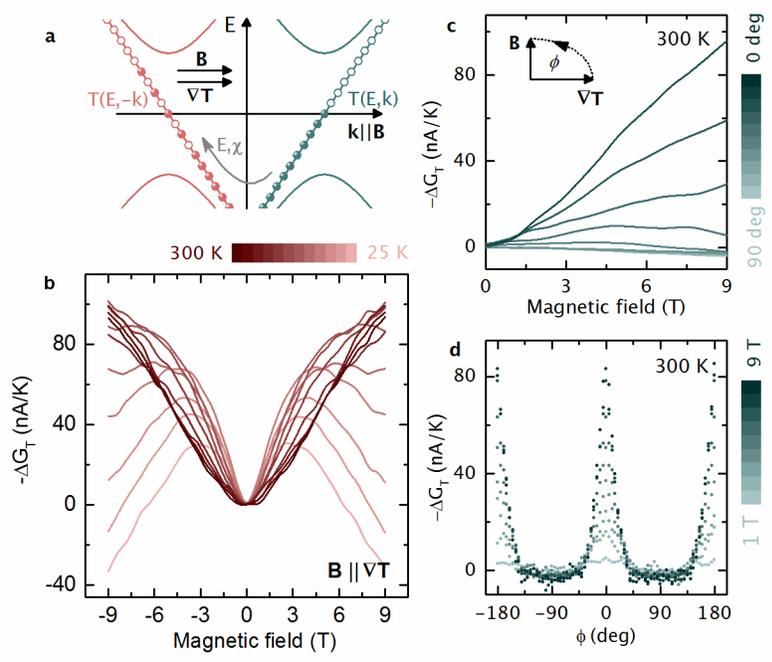

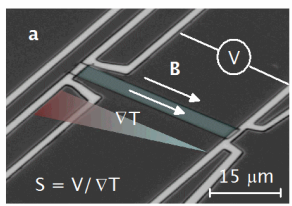
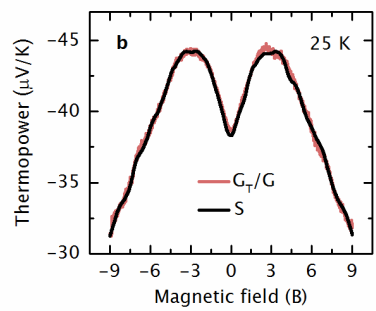

# Supplemental Information for the Manuscript

Experimental signatures of the mixed axial-gravitational anomaly in the Weyl semimetal NbP

*Johannes Gooth, Anna Corinna Niemann, Tobias Meng, Adolfo G. Grushin, Karl Landsteiner, Bernd Gotsmann, Fabian Menges, Marcus Schmidt, Chandra Shekhar, Vicky Süß, Ruben Hühne, Bernd Rellinghaus, Claudia Felser, Binghai Yan, Kornelius Nielsch*

correspondence to: jog@zurich.ibm.com

**This PDF file includes:**

**Author contributions**

J.G. conceived the original idea for the study. M.S., C.S., and V.S. synthesized the single-crystal bulk samples. R.H. characterized the crystal structure. B.R. supervised the micro-ribbon definition and the compositional analysis. A.N. fabricated the samples. J.G. carried out the thermoelectric transport measurements with the help of A.N.. J.G., A.N., F.M., B.G, T.M., and A.G.G. analyzed the data. B.G., C.F., B.Y. and K.N. supervised the project. A. G. G., T. M. and K. L. provided the theoretical background of the work. All authors contributed to the interpretation of the data and to the writing of the manuscript.

**Materials and Methods**

Micro-ribbon fabrication

High-quality single bulk crystals of NbP are grown via a chemical vapor transport reaction using an iodine transport agent. A polycrystalline powder of NbP is synthesized by direct reaction of niobium (Chempur 99.9%) and red phosphorus (Heraeus 99.999%) within an evacuated fused silica tube for 48 h at 800 °C. The growth of bulk single crystals of NbP is then initialized from this powder by chemical vapor transport in a temperature gradient, starting from 850 °C (source) to 950 °C (sink) and a transport agent with a concentration of 13.5 mg cm$^{-3}$ iodine (Alfa Aesar 99.998%).

Subsequently, NbP micro-ribbons are cut out from the bulk crystals (Figure S1a,b) using Ga focused ion beam (FIB) etching ($U = 30$ kV; $I = 65$ nA - 80 pA). The samples are prepared such that their longitudinal direction coincides with the [100] crystal axis of NbP. FIB etching caused *in-situ* Ga doping of the ribbons. The final composition of the samples is analysed by SEM-EDX, yielding 53 % Nb, 45 % P and 2 % Ga (Figure S1c). We study ribbons 50 µm times 2.5 µm times 500 nm in size. The dimensions of the ribbons are obtained from SEM. The high aspect ratio of our samples ensures the suppression of jet currents during the transport measurements. The single-crystallinity of the ribbons is evident from the X-ray diffraction pattern shown in Figure S1d.

Device fabrication

After cutting, the NbP micro-ribbons are directly transferred onto Si/SiO$_2$ chips in a solvent-less approach, using a micromanipulator. A soft mask for electrical contacts to the micro-ribbon is defined via laser-beam lithography (customized µPG system). A double layer of photoresist is spin-coated and baked (first photoresist: LOR 3B, spin-coated at 3500 rpm for 45 s, baked at 180 °C for 250 s; second photoresist: MAP 1205, spin-coated at 3500 rpm for 30 s, baked for 30 s at 100 °C), prior to laser exposure and development for 40 s in MAD 331 solution. Sputter deposition of Ti/Pt (10 nm/400 nm) is performed after 5 min of argon sputter cleaning. Titanium serves as adhesion promoter and diffusion barrier for the Pt. Finally, a lift-off is performed, 1 h at 80 °C in Remover 1165. An optical micrograph of a final device is shown in Figure S2.

Thermoelectric measurements

Thermoelectric measurements are performed in temperature-variable cryostat (Dynacool, Quantum Design) in vacuum. The cryostat is equipped with a ±9 T superconducting magnet. After fabrication, the micro-ribbon devices are wire bonded and mounted on a sample holder that allows rotation in angles from -10 to 370 deg. We specifically investigate three thermoelectric transport parameters of the NbP micro-ribbons: the electrical conductance $G = J/V$, the thermoelectric conductance $G_T = J/\nabla T$, and the thermopower $S = - V_{th}/\nabla T$, where $J$ denotes the electrical current, $V$ the voltage bias, $\nabla T$ a temperature gradient, and $V_{th}$ the voltage response to $\nabla T$. All transport coefficients are measured in the linear response regime as a function of the cryostat base temperature $T$, magnetic field **B** and rotation angle $\phi$, which is defined with respect to the transport direction along the longitudinal axis of the samples.

Electrical conductance measurements in a two-probe configuration are carried out under isothermal conditions with dc bias voltages up to $V = 1$ mV applied with a Yokogawa voltage source and a 100-times voltage divider across the length of the ribbon. Our results are independent of contact iteration. The responding current signal $J$ is enhanced using a current preamplifier (Stanford Research, model SR570) with an input impedance of 1 Ω. Contact lines across the full width of the sample ensure a homogeneous electric field distribution in the ribbon. Temperature-dependent $J$-$V$ measurements (Figure S3) reveal ohmic contacts. The magneto-conductance is measured by sweeping the magnetic field with 5 mT/s.

Thermoelectric conductance measurements are performed with the same current recording setup, but, crucially, without electric field imposed at the sample. Instead, the on-chip Joule heater line near the end of the micro-ribbon is used to generate a temperature gradient $\nabla T$ along the length of the micro-ribbon. $\nabla T$ is measured by resistive thermometry, using two metal four-probe thermometer lines located at the ends of the micro-ribbons (Figure S2). The thermometers are driven by lock-in amplifiers (Stanford Research, model SR830) with a 500 nA ac bias current at distinct frequencies (< 15 Hz) to prevent crosstalk. The thermometers are calibrated under isothermal conditions against the base temperature of the cryostat (Figure S4), from which $T(R)$ is determined. The heating voltage $V_H$ applied across a 2 k$\Omega$ shunt resistance at the heater line (Figure S5) is chosen such that a linear response of the current $J$ to the temperature gradient $J \propto \nabla T$ is ensured (Figure S6). $G_T$ is then obtained from linear fits of $J$ to versus the corresponding $\nabla T$ at fixed base temperatures, and exhibits temperature differences of up to 350 mK. The magneto-thermoelectric conductance is measured at a fixed heating voltage of 12 V, sweeping the magnetic field with 5 mT/s. $\nabla T$ showed no dependence on the magnetic field. The field sweeping rate is chosen such that the results are independent of it.

To obtain $S$, the current measurement setup is removed and replaced by a nanovoltmeter (Keithley, model 182A) to measure the open circuit voltage $V_{th}$ of the NbP micro-ribbons in response to $\nabla T$. Choosing the same $V_H$ as in the thermoelectric conductance measurements, ensures a linear response of $V_{th}$ to $\nabla T$ (Figure S7). $S$ is then extracted from linear fits of $V_{th}$ versus $\nabla T$.

**Temperature dependence of the transport coefficients at zero magnetic field**

In the absence of an applied magnetic field, the electrical conductance exhibits a non-metallic $G(T)$ dependence with a negative temperature coefficient of resistance, see Figure S8. Furthermore, without magnetic and electric fields applied along the sample, the measurement of thermoelectric conductance versus temperature changes sign at approximately 250 K (Figure S9). Below 250 K the transport in the NbP micro-ribbons is dominated by conduction-band electrons, indicated by a negative sign of the thermoelectric conductance in the $G_T$ versus $T$ plot. Because of the zero bandgap, holes are thermally excited in parallel, governing the transport at higher temperatures (positive $G_T$). The same generic features are observed in $S(T)$ (Figure S10).

**Additional data**

We provide additional data, including the original longitudinal $G(B)$ (Fig S11), $G_T(B)$ (Figure S12) and $S(B)$ (Figure S13) at selected $T$.

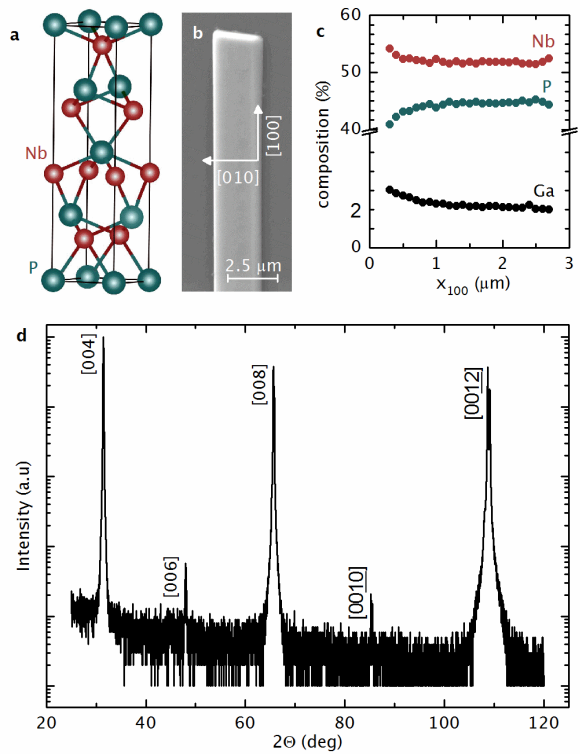

**Figure S1.**

Material analysis of the NbP micro-ribbon. (**a**) Sketch of the structure of the NbP crystal. (**b**) SEM image of a NbP micro-ribbon before device processing. The longitudinal direction of the ribbon corresponds to the [100] axis of the crystal. (**c**) Spatial composition of a NbP micro-ribbon, measured from the top along [100], using SEM-EDX, reveals an average of 53 % Nb, 45 % P and 2 % Ga. (**d**) XRD spectrum of the NbP at room temperature (Cu-K$_\alpha$ radiation).

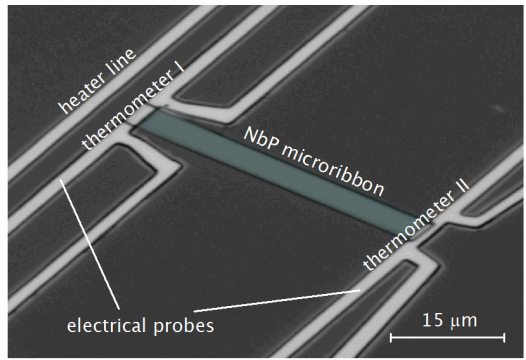

**Figure S2.**

Optical micrograph of a measurement device. The NbP micro-ribbon is placed between two four-probe thermometers, which also serve as electrical probes. The electrically insulated heater line close to one end of the sample creates a temperature gradient along the length of the sample.

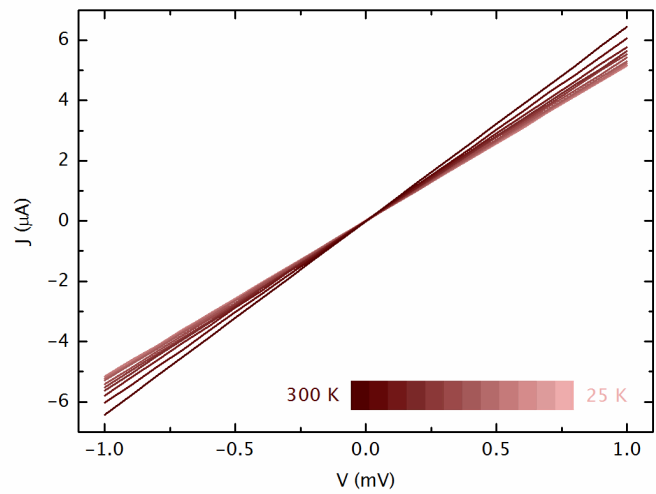

**Figure S3.**

Isothermal ($\nabla T = 0$ K) Current ($J$)-voltage($V$) characteristic of the NbP micro-ribbon at selected temperatures and zero magnetic field (**B** = 0 T). The linearity of the curves reveals ohmic electrical contacts.

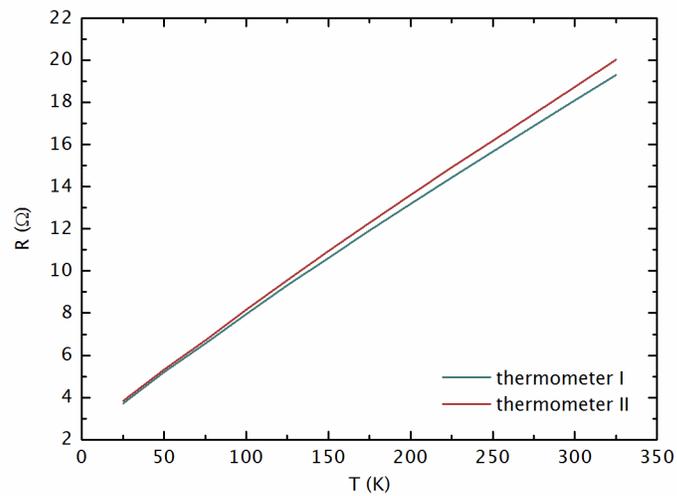

**Figure S4.**

Thermometer calibration. Resistance *R* versus base temperature *T* of the cryostat, measured at isothermal conditions.

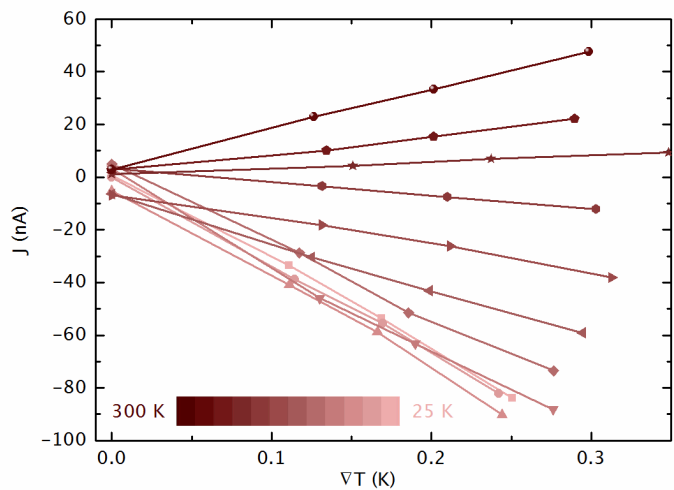

**Figure S5**

Temperature gradient ∇T along the sample as a function of the square of the heating voltage $V_H$, which is proportional to the heating power.

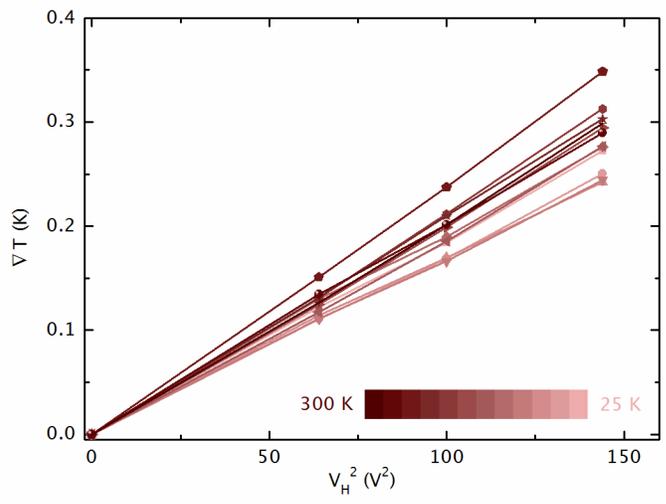

**Figure S6.**

Linear response of the thermoelectric current $J$ to the temperature gradient $\nabla T$. From the slope of the linear fits, the thermoconductance $G_T = J/\nabla T$ is determined. The error bars in Figure 1e in the main text are the uncertainties obtained from the linear fits.

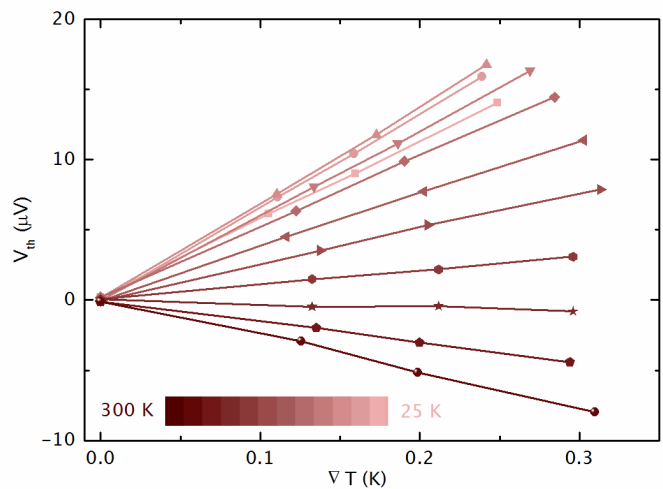

**Figure S7.**

Linear response of the thermovoltage $V$ to the temperature gradient $\nabla T$. The thermopower $S = - V_{th}/\nabla T$ is determined from linear fits of the data.

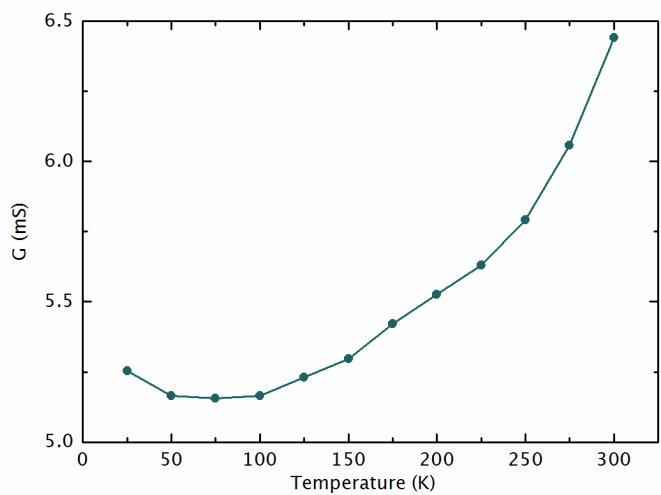

**Figure S8.**

Electrical conductance at zero magnetic field (**B** = 0 T) in isothermal conditions ($\nabla T = 0$ K) as a function of the base temperature. Values of $G$ are obtained from the slope of the linear fits of the data given in Figure S3.

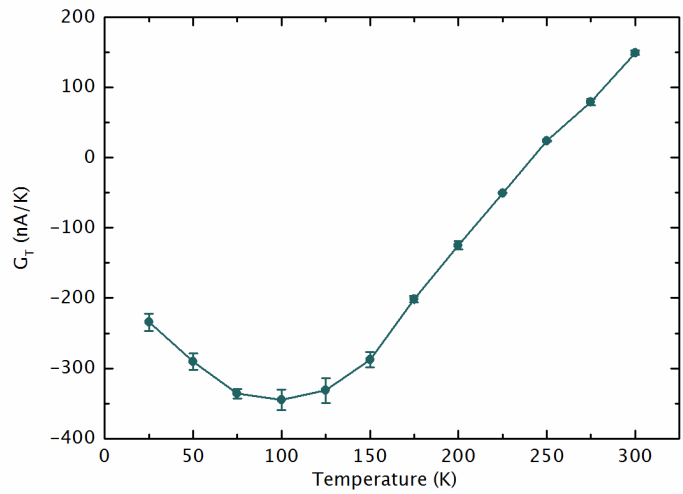

**Figure S9.**

Thermoelectric conductance at zero magnetic field (**B** = 0 T) with no electric field imposed (**E** = 0) as a function of the base temperature. Values of $G_T$ are obtained from the slope of the linear fits of the data shown in Figure S6. The error shown here is the fit uncertainty of the slope.

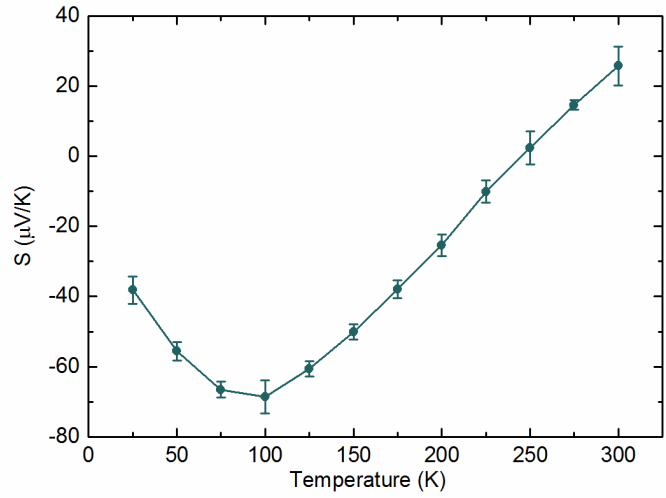

**Figure S10.**

Themopower at zero magnetic field (**B** = 0 T) as a function of the base temperature. Values of *S* are obtained from the slope of the linear fits of the data shown in Figure S7. The error shown here is the fit uncertainty of the slope.

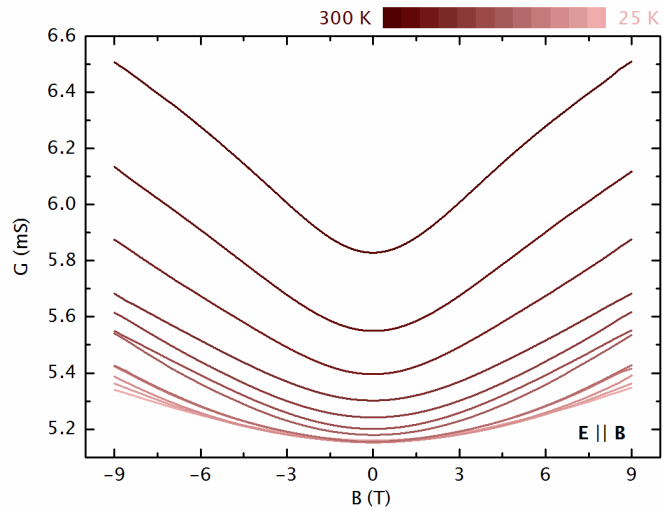

**Figure S11.**

Longitudinal magneto-conductance (**E**‖**B**) at selected base temperatures.

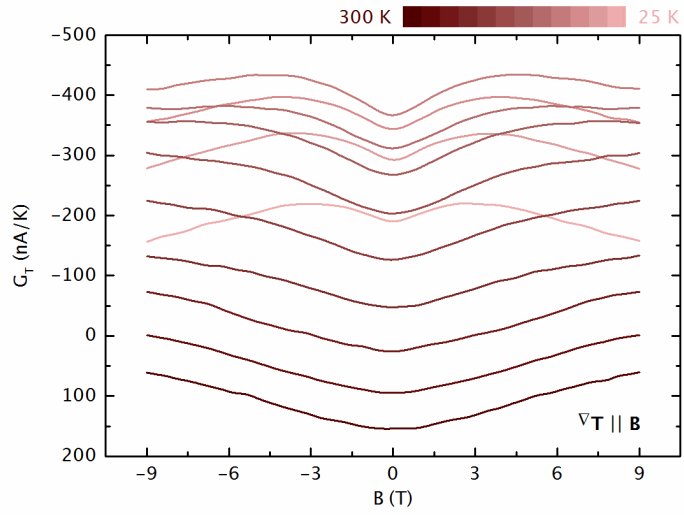

**Figure S12.**

Longitudinal magneto-thermoelectric conductance ($\nabla T \parallel \mathbf{B}$) at selected base temperatures.

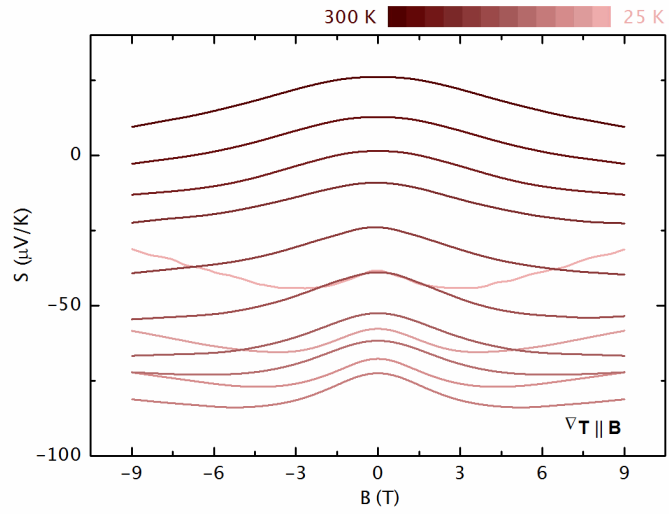

**Figure S13.**

Longitudinal magneto-thermopower ($\nabla T \parallel \mathbf{B}$) at selected base temperatures.

# 1 Connection between the mixed axial-gravitational anomaly and thermal transport

**The mixed axial-gravitational anomaly**  In this section we collect for completeness established quantum field theory and hydrodynamic theory results for chiral Weyl fermions which support the interpretation that the measurements presented in the main text stem from a mixed axial-gravitational anomaly. We start by noting that, unlike pure gravitational anomalies which are only allowed in space-time dimension $D = 4k - 2$, mixed axial-gravitational anomalies are allowed in $D = 3 + 1$ relevant for Weyl semimetals (1). In particular, we are interested in the abelian version of the mixed axial-gravitational anomaly which is expressed via the non-conservation of axial current $j_A^\mu$ in a curved space time characterized by the Riemann tensor $R^\alpha_{\beta\mu\nu}$. Following the notation of Ref. 2 it is possible to write the conservation law for this current as

$$\partial_\mu j_A^\mu = \frac{d_{AVV}}{32\pi^2} \varepsilon^{\mu\nu\rho\sigma} F^V_{\mu\nu} F^V_{\rho\sigma} + \frac{b_A}{768\pi^2} \varepsilon^{\mu\nu\rho\sigma} R^\alpha_{\beta\mu\nu} R^\beta_{\alpha\rho\sigma}, \qquad (1)$$

where the labels $A$ and $V$ denote tensors built out of axial and vector fields respectively and $d_{AVV}$ and $b_A$ are numerical coefficients defined shortly. The first term on the r.h.s. of Eq. (1) represents the non-conservation of axial current due to the presence of external non-orthogonal electric and magnetic field. It involves the electromagnetic (vector) field strength $F^V_{\mu\nu} = \partial_\mu A^V_\nu - \partial_\nu A^V_\mu$ through the $U(1)$ vector gauge field $A^V_\mu$ where $\mu = 0, 1, 2, 3$. Its coefficient is set by the chiral anomaly coefficient $d_{abc}$ where $a, b, c = \{A, V\}$; when $d_{abc} \neq 0$ the chiral anomaly is present. For the abelian case, of interest here, it is simply determined by the difference between a triple product of charges of right (R) and left (L) chiralities:

$$d_{abc} = \sum_r (q_a^r q_b^r q_c^r) - \sum_l (q_a^l q_b^l q_c^l), \qquad (2)$$

where $q^r_{V,A} = (1,1)$ and $q^l_{V,A} = (1,-1)$. Therefore $d_{AVV} = 2$ for a pair of chiral fermions.

The second term on the r.h.s. of Eq. (1) is the mixed axial-gravitational anomaly contribution to the non-conservation of chiral current. In the abelian case its coefficient is given by

$$b_a = \sum_r q^r_a - \sum_l q^l_a. \qquad (3)$$

with $a = A, V$. If $b_A \neq 0$ the mixed axial-gravitational anomaly is present $(1,3,4)$ and $b_A = 2$ for each pair of chiral fermions.

**Connection with thermal transport**   Given the form of the gravitational term in Eq. (1), it is natural to ask how it is possible to detect its presence in a flat space-time where $R^\alpha_{\beta\mu\nu} = 0$. The key and subtle observation used in the main text is that the temperature dependence of linear transport coefficients for systems of Weyl fermions depends on $a_g$, even in flat-space time. This conclusion can be reached from either the standard Kubo formalism (6), holography (7), hydrodyanamic theory (8) or arguments based on a global version of the axial-gravitational anomaly (9).

Although hydroydnamic (10) and Kubo approaches lead to consistent predictions for the thermoelectric coefficient presented in this work, the former relies on interactions being the dominant scattering mechanism. Because the scattering in experimental samples seems to be mostly dominated by impurity scattering, we believe the Kubo approach to be more suitable for the description of our data.

In what follows we sketch a simple derivation that justifies the functional form of the PMTC used in the main text. We specialise to a single Weyl cone, in which case the anomaly coefficients are $d_\chi = b_\chi = \pm 1$, and we define $a_\chi = d_\chi/(4\pi^2)$ and $a_g = b_\chi/24$. Our calculation relies on the Kubo formalism, and while the low field predictions obtained by this treatment are consistent with the ones from Boltzmann kinetic theory (11–14), key advantages of the Kubo formalism as compared to the Boltzmann approach are that it tracks the relationship to the mixed axial-gravitational anomaly in a transparent way, and that it is directly applicable to large magnetic fields. We start with the equations that describe particle and energy conservation for a single Weyl fermion,

$$\dot{\rho} + \vec{\nabla} \vec{J}_\rho = a_\chi \vec{E} \cdot \vec{B}, \tag{4}$$

$$\dot{\epsilon} + \vec{\nabla} \vec{J}_\epsilon = \vec{J}_\rho \cdot \vec{E}, \tag{5}$$

where $\rho$ is the electronic density, and $\epsilon$ denotes the energy density. Because the system lives in flat spacetime, only the chiral anomaly enters these equations. We furthermore note that the energy conservation equation (5) includes a term describing the work performed by the electric field.

We are interested in the magnetic field dependent contribution to the energy and charge current which will ultimately determine the PMTC. As discussed in Ref. (2), the standard Kubo formalism for Weyl fermions leads to

$$\vec{J}_\rho = a_\chi \mu \vec{B}, \tag{6}$$

$$\vec{J}_\epsilon = \left(\frac{a_\chi}{2}\mu^2 + a_g T^2\right)\vec{B}. \tag{7}$$

Eq. (6) describes the chiral magnetic effect for a single Weyl fermion at chemical potential $\mu$, which depends on the chiral anomaly coefficient $a_\chi$. The energy current given in Eq. (7) is composed of two terms. The first one describes that the directed flow of particles in a system with a chiral anomaly leads to an energy current that is simply due to the energy associated with each electronic state. This term is normalized such that the energy current vanishes in the vacuum state, which corresponds to $\mu = 0$. The second term, which describes the thermal contribution of interest to us, has been recently pointed out to be a consequence of mixed axial-gravitational anomaly (6–9). It is hence governed by the coefficient $a_g$, which establishes a link between the existence of a thermal contribution in $\vec{J}_\epsilon$ and the gravitational contribution to Eq. (1).

We now turn to the thermoelectric transport. To this end, we assume a finite magnetic field $\vec{B}$ and compute the anomalous response due to a gradient in temperature $\vec{\nabla}T$ and an electric field $\vec{E}$ (all of which are assumed to be spatially homogeneous) in the linear response approximation. With these assumptions we can insert Eqs. (6) and (7) in Eqs. (4) and (5) to obtain

$$\dot{\epsilon} = a_\chi \mu \vec{E}.\vec{B} - 2a_g T \vec{\nabla}T.\vec{B}, \tag{8}$$

$$\dot{\rho} = a_\chi \vec{E}.\vec{B}. \tag{9}$$

The first term on the right hand side of (8) is the work performed by the electric field to the background chiral magnetic effect current. If there are two nodes of opposite chirality, this term encodes that each particle pumped between the Weyl nodes in parallel electric and magnetic fields also transfers its energy from one node to the other. The second term in the right hand side of (8) shows that, in a completely analogous way, a temperature gradient parallel to the magnetic field leads to energy pumping between Weyl cones due to the mixed axial-gravitational anomaly. In a Weyl semimetal intervalley scattering will stop this pumping on a timescale $\tau$, the intervalley scattering time (in a strongly interacting electron fluid the energies of different Weyl cones could in principle equilibrate on a shorter timescale due to electron-electron interactions - this, however, is not the case in the present experiment). The steady state is accounted for by the replacement $(\dot{\epsilon}, \dot{\rho}) \to \frac{1}{\tau}(\delta\epsilon, \delta\rho)$. At this point it is convenient to introduce the matrix

$$\Xi = \begin{pmatrix} \frac{\partial \epsilon}{\partial T} & \frac{\partial \epsilon}{\partial \mu} \\ \frac{\partial \rho}{\partial T} & \frac{\partial \rho}{\partial \mu} \end{pmatrix} \tag{10}$$

such that

$$\begin{pmatrix} \delta\epsilon \\ \delta\rho \end{pmatrix} = \Xi \cdot \begin{pmatrix} \delta T \\ \delta\mu \end{pmatrix} \tag{11}$$

Now we write Eqs. (8) and (9) as

$$\Xi \cdot \begin{pmatrix} \delta T \\ \delta\mu \end{pmatrix} = \tau \begin{pmatrix} -2a_g T & a_\chi \mu \\ 0 & a_\chi \end{pmatrix} \begin{pmatrix} \vec{\nabla} T \\ \vec{E} \end{pmatrix} \cdot \vec{B} \tag{12}$$

From Eqs. (6) and (7), one furthermore obtains that the $k$-th component of the energy and particle currents are given by

$$\begin{pmatrix} \delta J_{\epsilon,i} \\ \delta J_{\rho,i} \end{pmatrix} = \begin{pmatrix} 2a_g T & a_\chi \mu \\ 0 & a_\chi \end{pmatrix} \cdot \begin{pmatrix} \delta T \\ \delta\mu \end{pmatrix} B_i \tag{13}$$

which together with Eq. (12) defines the response tensor $\kappa_{ij}$ as

$$\kappa_{ij} = \begin{pmatrix} 2a_g T & a_\chi \mu \\ 0 & a_\chi \end{pmatrix} \cdot \Xi^{-1} \cdot \begin{pmatrix} -2a_g T & a_\chi \mu \\ 0 & a_\chi \end{pmatrix} \tau B_i B_j \tag{14}$$

The anomaly induced response in the current is thus given by

$$J_i = G_T^{ij} \nabla_j T + G^{ij} E_j, \tag{15}$$

where the conductivity tensors have components

$$G^{ij} = \tau \frac{a_\chi^2}{\det(\Xi)} \left( \frac{\partial \epsilon}{\partial T} - \mu \frac{\partial \rho}{\partial T} \right) B_i B_j, \quad (16)$$

$$G_T^{ij} = \tau \frac{2 a_\chi a_g T}{\det(\Xi)} \frac{\partial \rho}{\partial T} B_i B_j. \quad (17)$$

To explain our experimental measurements, we now specialize on the regimes of either low or high magnetic field, assuming a homogenous Fermi velocity of $v_F$ (we use units such that $e = 1$). In the first case, the system behaves as a gas of free Weyl fermions with an energy density of

$$\epsilon = \frac{1}{8\pi^2 v_F^3} \left( \mu^4 + 2\pi^2 T^2 \mu^2 + \frac{7}{15} \pi^4 T^4 \right) \quad (18)$$

(see for instance Ref. (15)). For large magnetic fields, the system splits into degenerate one-dimensional systems, the number of which is set by the Landau level degeneracy. In this case the energy density reads

$$\epsilon_B = \frac{|\mathbf{B}|}{v_F} \left( \frac{\mu^2}{8\pi^2} + \frac{T^2}{24} \right). \quad (19)$$

For small and large magnetic fields, the particle densities are given respectively by

$$\rho = \frac{\mu^3 + \pi^2 T^2 \mu}{6\pi^2 v_F^3}, \quad (20)$$

$$\rho_B = \frac{|\mathbf{B}|}{4\pi^2 v_F} \mu. \quad (21)$$

Combining the above allows us to obtain $G_T^{ij}$ and $G_T^{ij}$ in different magnetic field regimes. For low magnetic fields, using Eqs. (18) and (20) in (16) and (17) we obtain

$$G^{ij} = \frac{6a_\chi^2 \pi^2 v_F^3 (7\pi^2 T^2 + 5\mu^2)}{7\pi^4 T^4 + 6\pi^2 T^2 \mu^2 + 15\mu^4} \tau B_i B_j, \quad (22)$$

$$G_T^{ij} = \frac{120 a_\chi a_g \pi^2 T \mu v_F^3}{7\pi^4 T^4 + 6\pi^2 T^2 \mu^2 + 15\mu^4} \tau B_i B_j \quad (23)$$

while for large magnetic fields Eqs. (19) and (21) in (16) and (17) lead to

$$G^{ij} = 4a_\chi^2 \pi^2 v_F \tau \frac{B_i B_j}{|\mathbf{B}|} \tag{24}$$

$$G_T^{ij} = 0. \tag{25}$$

By adding a Drude contribution and summing over the contributions from the different nodes, we obtain the functional form used in the main text.

Eq. (17) and (23) establish a link between the presence of the chiral and mixed axial-gravitational anomaly, represented by $a_\chi \neq 0$ and $a_g \neq 0$, and the enhancement of the thermoelectric response function of Weyl fermions in flat space-time due to a magnetic field applied parallel to a thermal gradient. In particular, if the axial-gravitational anomaly is absent there is no thermal transport. Since we probe the presence of the chiral anomaly ($a_\chi \neq 0$) independently by measuring a positive magneto-conductance, our results indicate that $a_g \neq 0$. This is the main finding of our work.

Physically, the effect can be summarized as follows. The thermal gradient increases (decreases) the energy density of left (right) movers, which leads to an effective chiral imbalance that generates a current in the direction of the magnetic field via the chiral magnetic effect. Finally, as mentioned above and in the main text, the functional dependence used in here is consistent with more conventional semiclassical treatments based on the Boltzmann equation (11–14), that however do not explicitly track the anomalous origin of this correction, and cannot directly access the high magnetic field limit.